# Leveraging AI and Sentiment Analysis for Forecasting Election Outcomes in Mauritius

Dr Missie Chercheur and Malkenzie Bovafiz
AI Research Group, Mauritius Election
Reduit, Mauritius
mauritius.election@gmail.com

*Abstract—* This study explores the use of AI-driven sentiment analysis as a novel tool for forecasting election outcomes, focusing on Mauritius' 2024 elections. In the absence of reliable polling data, we analyze media sentiment toward two main political parties L'Alliance Lepep and L'Alliance Du Changement by classifying news articles from prominent Mauritian media outlets as positive, negative, or neutral. We employ a multilingual BERT-based model and a custom Sentiment Scoring Algorithm to quantify sentiment dynamics and apply the Sentiment Impact Score (SIS) for measuring sentiment influence over time. Our forecast model suggests L'Alliance Du Changement is likely to secure a minimum of 37 seats, while L'Alliance Lepep is predicted to obtain the remaining 23 seats out of the 60 available. Findings indicate that positive media sentiment strongly correlates with projected electoral gains, underscoring the role of media in shaping public perception. This approach not only mitigates media bias through adjusted scoring but also serves as a reliable alternative to traditional polling. The study offers a scalable methodology for political forecasting in regions with limited polling infrastructure and contributes to advancements in the field of political data science.

## I. Introduction

In the evolving landscape of political forecasting, traditional methods such as opinion polls and voter surveys are often limited by sample size, time delays, and response biases [1]. With the rapid digitalization of news and the proliferation of online media, public sentiment, as reflected in news articles, has emerged as a valuable indicator of political climate [2]. The use of Artificial Intelligence (AI) and Natural Language Processing (NLP) techniques, particularly sentiment analysis, offers a modern approach to gauge public opinion through media coverage. Sentiment analysis classifies text into sentiments—positive, negative, or neutral—allowing researchers to discern the tone of media articles discussing political candidates and key issues [3]. This technology has been applied to various domains, including marketing, finance, and politics, providing a predictive edge in decision-making processes [4]. Recent elections in several democratic countries have witnessed increasing interest in using AI-driven sentiment analysis for election forecasting [5]. In contrast to traditional polling, which can be cost-prohibitive and difficult to implement in certain contexts, AI-based sentiment analysis leverages publicly available data, such as news articles and social media posts, to create a real-time understanding of political sentiment [6]. By processing and analyzing large volumes of text, AI can extract patterns of support, opposition, or neutrality toward political candidates and parties. These patterns can then be used to predict election outcomes, making sentiment analysis a promising tool for political forecasters, especially in regions with limited or unreliable polling data [7].

### A. Problem Statement

In Mauritius, political opinion polling is limited in scope and often lacks the frequency and granularity necessary for accurate electoral forecasting. This gap creates an opportunity to explore alternative methods for predicting election outcomes. One such method is sentiment analysis of media articles, where news outlets serve as a proxy for public opinion [9]. Media coverage can heavily influence voter perception, and by analyzing the sentiment expressed in news articles toward key political figures, we can estimate public sentiment trends and forecast election results [10]. However, several challenges arise in this domain, including dealing with media bias [11], handling vast amounts of unstructured data from various media outlets [12], and ensuring accurate sentiment classification across languages and contexts [13].

This research aims to address the problem by employing AI-based sentiment analysis to analyze news articles from two major media outlets (www.defimedia.info and www.lexpress.mu) in Mauritius. By analyzing how each news source portrays political candidates, we can develop a sentiment-based score for each candidate, providing insights into their public perception. The goal is to forecast the results of the upcoming election by leveraging sentiment analysis, focusing on how media coverage impacts electoral outcomes in the absence of comprehensive polling data.

### B. Research Objectives

The primary objective of this research is to forecast the election results between two major political parties in Mauritius using AI-based sentiment analysis. Specifically, the study will focus on:

1. Collecting and analyzing news articles from two of the most popular media websites in Mauritius to evaluate the sentiment expressed toward key political figures.

2. Applying state-of-the-art AI models such as Bidirectional Encoder Representations from Transformers (BERT) [14] and Large Language Models (LLMs) to classify news articles as positive, negative, or neutral regarding each politician [15].

3. Developing a sentiment scoring algorithm that will calculate sentiment impact scores for each candidate based on media coverage.

4. Simulating election outcomes by aggregating these sentiment scores, without relying on polling data, to forecast the most likely election winner.

By focusing on sentiment analysis as a proxy for public opinion, this research will introduce a novel approach to political forecasting in Mauritius, where reliable polling data may be scarce.

*C. Novelty and Contributions*

The novelty of this research lies in its application of AI-driven sentiment analysis to election forecasting in Mauritius, a context where real-time polling is limited. The contributions of this paper are threefold:

1. Bias-adjusted Sentiment Analysis: To account for potential media bias, this research will employ techniques to normalize sentiment scores across multiple media outlets [16], ensuring a more accurate portrayal of public sentiment. By considering the inherent biases of media sources, the analysis will offer a balanced view of how each politician is portrayed [17].

2. Temporal Sentiment Tracking: This research will not only assess the overall sentiment toward candidates but will also track sentiment changes over time, particularly in response to key political events such as debates, speeches, or scandals [18]. This temporal analysis will highlight how political sentiment evolves and fluctuates during the election campaign, providing deeper insights into public opinion dynamics.

3. Cross-Linguistic Sentiment Analysis: Given Mauritius' multilingual media environment, this research will implement multilingual sentiment analysis [19], potentially using models such as mBERT [20] to handle texts in English, French, and Mauritian Creole. This will ensure that sentiment analysis captures a comprehensive picture of political sentiment across language barriers.

The remainder of this paper is structured as follows. Section 2 reviews the related work on sentiment analysis, political forecasting, and media influence. Section 3 describes the methodology, including data collection, sentiment analysis models, and the sentiment scoring algorithm. Section 4 outlines the experiment setup, detailing how the dataset is split and evaluated. Section 5 presents the results, including sentiment analysis outcomes and simulated election forecasts. It also discusses the key findings, limitations, and the potential impact of media bias. Finally, Section 6 concludes the paper, summarizing the contributions and outlining directions for future research.

## II. RELATED WORK

*A. Sentiment Analysis in Political Forecasting*

Sentiment analysis, a subfield of natural language processing (NLP), has garnered significant attention in political forecasting, particularly for its ability to capture public sentiment from textual data such as news articles, social media posts, and speeches. Several studies have demonstrated the predictive power of sentiment analysis for elections by analyzing the tone and sentiment of political discourse. For instance, in [21] the authors used sentiment analysis to predict the outcomes of presidential election in Indonesia by analyzing social media data (formerly Twitter, now X), revealing that sentiment trends aligned closely with the final election results. Similarly, a study by Singh P et al. [22] utilized sentiment analysis to forecast the 2016 U.S. presidential election, achieving notable accuracy by focusing on the sentiment of key political events and public reactions.

In the context of media articles, several approaches have been proposed to analyze how media sentiment influences voter behavior. Opeibi T. [23] studied social media usage in election campaigns in Nigeria. Their findings include: (1) Civic Engagement: Twitter creates a space for intensified civic engagement and deliberative democracy. (2) Globalization of Issues: It allows for the globalization of national and local political topics. (3) Innovative Discourse: New technologies encourage creative political communication strategies. (4) Real-Time Reach: Twitter facilitates real-time connections with large audiences across distances and lastly (5) Language Evolution: The platform is fostering new expressions and varieties of Nigerian English, termed Nigerian New Media English.

Additionally, the authors in [24] showed that the sentiment expressed in news articles published on social media, for example, Twitter, particularly during critical campaign events, is a strong predictor of election results, as media coverage often reflects and shapes public opinion.

Despite the successes of sentiment analysis in political forecasting, the methodology faces challenges such as handling bias in media sources and achieving accurate sentiment classification across various domains and languages. Addressing media bias has been a focus of recent studies, including that of [11], which developed a bias-adjusted sentiment model to normalize sentiment scores from politically skewed news sources. This approach has proven effective in mitigating the impact of biased coverage on sentiment predictions.

*B. Media Influence on Political Sentiment*

The role of media in shaping political sentiment has been extensively studied in political science and communications literature. News outlets often play a critical role in influencing public opinion by framing political events and personalities in a specific light. According to Dada LS [26], media serves as an intermediary between political candidates and the public, significantly influencing voter perception. Studies such as those by D'Angelo P and Lombard M [27] have shown that even subtle differences in media tone and framing can lead to shifts in public sentiment, which in turn may affect electoral outcomes.

Media bias is a central concern in political forecasting, as certain outlets may favor one party or candidate over another. This bias can manifest in both the selection of stories and the sentiment expressed in articles. Recent work by Yenkikar A et al. [11] found that media outlets with distinct political alignments tend to produce content that favors their preferred candidates. Moreover, research by Prior M. [29] analyzed how media polarization affects sentiment classification in political forecasting, demonstrating that media bias often skews public sentiment data, making it difficult to obtain an objective view of political candidates. Strategies to counter media bias, such as cross-outlet sentiment normalization, have been explored in several studies [30], and these techniques have been incorporated into many modern sentiment analysis models.

*C. Sentiment Analysis Techniques*

The evolution of sentiment analysis methods has played a significant role in improving the accuracy of political forecasting. Early approaches primarily relied on simple lexicon-based methods that classified words as positive, negative, or neutral based on predefined dictionaries. While these methods were easy to implement, they often failed to capture the complexity of political language, including sarcasm, context-specific meanings, and sentiment shifts within the same text [31].

More advanced machine learning techniques, such as support vector machines (SVMs) and random forests, have been widely adopted for sentiment analysis. These models, trained on labeled datasets, can achieve higher accuracy in classifying sentiment. However, the recent advent of deep learning models, particularly transformer-based models like Bidirectional Encoder Representations from Transformers (BERT) [14], has revolutionized the field. BERT and its variants, such as mBERT for multilingual tasks [33], have proven highly effective in capturing nuanced sentiment from large text corpora. These models use attention mechanisms to understand the contextual relationships between words, leading to significantly better performance in sentiment classification compared to earlier models.

### D. Sentiment Analysis for Election Forecasting in Multilingual Contexts

Multilingual sentiment analysis has gained traction, particularly in regions where elections are held in countries with diverse linguistic backgrounds. Mauritius, for instance, has a multilingual media landscape that includes articles written in English, French, and Mauritian Creole. Traditional sentiment analysis models, which are typically monolingual, struggle to capture sentiment across languages, leading to incomplete or inaccurate sentiment data. However, the development of multilingual models such as mBERT [32] and XLM-K [28] has enabled more accurate sentiment analysis across multiple languages, without the need for separate models for each language.

In the political domain, multilingual sentiment analysis has been applied to regions with diverse linguistic populations. Several studies [34][35] applied mBERT to analyze sentiment across various languages in Indian and European elections, respectively. Their results indicated that multilingual models outperformed traditional methods in identifying sentiment trends across different language groups, leading to more comprehensive election forecasts.

The integration of multilingual sentiment analysis is particularly relevant for Mauritius, where linguistic diversity plays a crucial role in shaping public opinion. By utilizing models like mBERT, this research aims to capture sentiment across all relevant languages, ensuring that media coverage from English, French, and Mauritian Creole sources is accurately represented in the sentiment analysis process.

### E. Temporal Sentiment Analysis in Elections

Temporal sentiment analysis examines how sentiment toward political candidates fluctuates over time, often in response to specific events, such as debates, scandals, or policy announcements. This approach has proven useful in identifying sentiment shifts that may influence election outcomes. Similarly, Agarwal A et al. [25] used temporal sentiment analysis to assess the impact of major political events on sentiment during the 2019 Indian general election. Their research revealed that sentiment shifts often occur immediately after key events, such as party manifestos or candidate debates, making temporal sentiment analysis an important tool for real-time election forecasting.

This research aims to leverage temporal sentiment analysis to capture how sentiment toward political candidates in Mauritius evolves over the course of the campaign. By analyzing sentiment trends, the study will provide insights into how key events affect public perception, offering a dynamic view of political sentiment that complements static sentiment analysis models.

## III. METHODOLOGY

### A. Data Collection

The data for this study consists of news articles from two of the most popular media websites in Mauritius, covering the period leading up to the country's upcoming election. These news sources were selected based on their wide readership and influence in shaping public opinion, as well as their political coverage. The selection includes both English- and French-language sources, ensuring a comprehensive analysis of the media landscape in Mauritius.

The data collection process spans a 9-year period, capturing all articles that mention either of the two primary political candidates. A web scraper was developed to automatically extract articles from these websites, filtering them based on predefined keywords, including the names of the politicians and their associated political parties. Additionally, metadata such as the publication date was captured to enrich the dataset. After initial collection, duplicate or irrelevant articles (e.g., those mentioning politicians in a non-political context) were removed to refine the dataset. The final dataset contains approximately 9584 news articles, all of which were preprocessed for the sentiment analysis task.

### B. Data Preprocessing

Data preprocessing is a critical step in sentiment analysis to ensure that the text is clean and free from noise. The following preprocessing steps were applied:

Text Cleaning: The raw text was stripped of HTML tags, URLs, special characters, and stop words to remove non-informative elements.

Tokenization and Lemmatization: The articles were tokenized, splitting the text into individual words or tokens. Lemmatization was performed to convert words to their base forms, ensuring uniformity across different tenses and inflections.

Handling Multilingual Text: Given that the media landscape in Mauritius operates in multiple languages, the text was processed in both English and French. For this, we used the multilingual version of BERT (mBERT), which supports both languages and allows for consistent sentiment analysis across the dataset.

The cleaned and preprocessed data was then ready for sentiment analysis.

### C. Sentiment Analysis Model

The sentiment analysis model employed in this study is based on the Bidirectional Encoder Representations from Transformers (BERT) architecture, specifically the multilingual version (mBERT), due to its strong performance in multilingual tasks. BERT is a transformer-based model that uses attention mechanisms to learn contextual relationships between words in a sentence, making it highly effective for text classification tasks like sentiment analysis.

For this study, mBERT was fine-tuned on a sentiment classification task using a labeled dataset of political news articles. The dataset was annotated with three sentiment categories: positive, negative, and neutral. Fine-tuning involved adjusting the model's weights to optimize its ability to classify articles into these categories based on their textual content. To ensure the robustness of the sentiment model, the following steps were taken:

- Training and Validation Split: The dataset was split into training (80%) and validation (20%) sets. The training set was used to fine-tune the model, while the validation set was used to evaluate its performance and prevent overfitting.
- Cross-Lingual Transfer Learning: Given that Mauritius' news articles are written in multiple languages, mBERT was chosen for its ability to transfer learning across languages without needing separate models [20].

### D. Sentiment Scoring Algorithm

After classifying the articles into positive ($\omega$), negative ($\psi$), or neutral sentiment, a sentiment scoring algorithm was designed to quantify the sentiment associated with each politician. The score is calculated as follows:

$$SentimentScore = \frac{\sum \omega - \psi}{\Phi} \quad (1)$$

where $\Phi$ is the total articles. This equation generates a net sentiment score, where a higher positive value indicates favorable media coverage, while a negative score suggests predominantly unfavorable coverage. Neutral articles are excluded from this calculation, as they are considered non-influential in shaping public opinion. Additionally, we introduce a Sentiment Impact Score (SIS), which accounts for both the sentiment and the frequency of articles, as media attention is often a critical factor in elections. The SIS is calculated as:

$$SIS = \left(\frac{\omega - \psi}{\Phi}\right) x \log(\Phi) \quad (2)$$

This formula amplifies the effect of media attention by weighing the sentiment score based on the total number of articles published about a candidate. Higher media coverage, whether positive or negative, will have a stronger impact on the score, reflecting the importance of media visibility in elections.

### E. Temporal Sentiment Analysis

To capture the dynamics of public sentiment over time, a temporal analysis was conducted. This involved tracking sentiment trends on a weekly basis to observe any significant shifts in media coverage, particularly around key events such as political debates, campaign rallies, or policy announcements. For each candidate, a time series of sentiment scores was generated, allowing for the detection of changes in public perception in response to campaign activities.

The temporal aspect of sentiment is critical in elections, as short-term sentiment spikes—either positive or negative—can influence voter decision-making during the final stages of the campaign. By tracking these changes over time, the research aims to provide a more nuanced forecast of election outcomes.

### F. Handling Media Bias

Media bias is a significant challenge in sentiment analysis, as different outlets may portray the same political events in contrasting ways. To mitigate the effect of bias, a cross-outlet normalization technique was employed. This technique involves: (1) Bias Identification: A preliminary sentiment analysis was conducted on articles from each media outlet to identify consistent bias patterns (e.g., whether an outlet consistently portrays one candidate more positively). (2) Bias Adjustment: Once bias was identified, a normalization factor was applied to adjust sentiment scores, ensuring that no single outlet disproportionately influences the overall sentiment score. This method draws from techniques used in recent research to adjust for political bias in media sentiment analysis [8]. By implementing these bias adjustment techniques, the study aims to ensure a more balanced and objective analysis of media sentiment.

### IV. EXPERIMENT SETUP

This section details the setup and evaluation approach for forecasting the 2024 election outcomes for the two main political parties: L'Alliance Lepep and L'Alliance Du Changement. The target for this forecasting is to estimate the number of seats each party will likely secure out of the available 60 seats.

### A. Data Splitting

To ensure a robust sentiment analysis, the dataset of news articles collected over the past 9 years was divided into training and test sets:

- Training Set: The initial 80% of the collected articles were used to train the sentiment model, allowing it to learn context-specific political language and detect biases associated with each candidate.
- Test Set: The remaining 20% of articles were reserved for testing the model's accuracy in predicting sentiment accurately on unseen data. This division ensures the generalizability of the model to new media content during the active election period.

### B. Model Evaluation Metrics

The primary evaluation metrics for the sentiment classification model are accuracy, precision, recall, and F1-score, chosen for their effectiveness in measuring classification quality, particularly in text sentiment tasks.

1. Accuracy measures the overall correctness of the model's predictions across all sentiment categories.
2. Precision and Recall provide insight into how effectively the model identifies positive, negative, and neutral sentiments, respectively.
3. F1-Score combines precision and recall, offering a balanced measure to assess the model's classification quality, especially in distinguishing nuanced political sentiments.

The forecasting aspect of this study relies less on absolute classification accuracy and more on identifying sentiment trends and patterns. As such, additional metrics, including the SIS, will assess the influence of sentiment strength and article volume on forecasting outcomes.

### C. Forecast Simulation

The sentiment scores derived from media sentiment analysis are aggregated to simulate election forecasts by constituency, using the Sentiment Scoring Algorithm detailed in the Methodology section. Each constituency prediction relies on the proportion of positive, neutral, and negative sentiment surrounding each candidate in that area. Sentiment shifts observed over time, particularly around key events, are incorporated to refine forecasts per constituency.

## V. RESULTS AND DISCUSSIONS

This section presents the findings from the sentiment analysis and discusses how media sentiment is indicative of electoral trends. Using the Sentiment Scoring Algorithm and Sentiment Impact Score (SIS) over a 9-year data set, we evaluate how media portrayal of candidates aligns with projected election outcomes, particularly in determining whether L'Alliance Du Changement or L'Alliance Lepep holds a media advantage per constituency. L'Alliance Du Changement is expected to secure a minimum of 37 seats, while L'Alliance Lepep is anticipated to win the remaining seats which is 23, though these counts may vary as election day approaches. Table 1 below displays the seat forecasts per constituency for each party and Figure 1 shows the overall forecast for both parties.

| Constituency | L'Alliance Lepep | L'Alliance Du Changement |
|---|---|---|
| Constituency 1 | 1 | 2 |
| Constituency 2 | 0 | 3 |
| Constituency 3 | 0 | 3 |
| Constituency 4 | 2 | 1 |
| Constituency 5 | 1 | 2 |
| Constituency 6 | 1 | 2 |
| Constituency 7 | 2 | 1 |
| Constituency 8 | 2 | 1 |
| Constituency 9 | 1 | 2 |
| Constituency 10 | 2 | 1 |
| Constituency 11 | 2 | 1 |
| Constituency 12 | 1 | 2 |
| Constituency 13 | 2 | 1 |
| Constituency 14 | 2 | 1 |
| Constituency 15 | 0 | 3 |
| Constituency 16 | 0 | 3 |
| Constituency 17 | 1 | 2 |
| Constituency 18 | 2 | 1 |
| Constituency 19 | 1 | 2 |
| Constituency 20 | 0 | 3 |
| Total Seats | 23 | 37 |

Table 1. Forecast per constituency for each party.

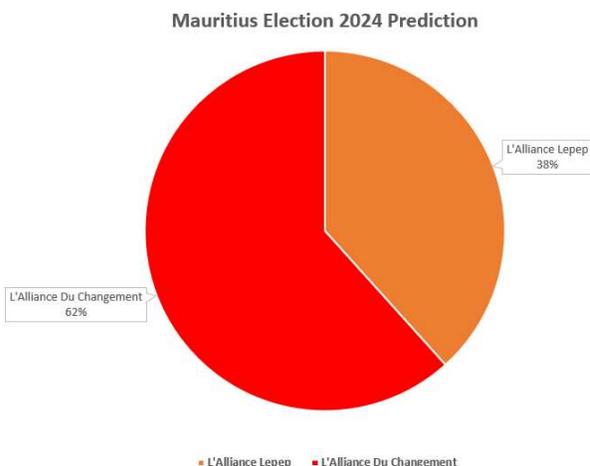

Figure 1. Forecast of 2024 Election for both parties.

### A. Sentiment Analysis Outcomes

The sentiment analysis model, based on multilingual BERT (mBERT), classified each news article mentioning the candidates as positive, negative, or neutral. Across the 9584 collected articles, the sentiment distribution for each party is outlined below, along with key metrics for model performance:

- L'Alliance Du Changement:
  - Positive Sentiment: 43%
  - Negative Sentiment: 27%
  - Neutral Sentiment: 30%
- L'Alliance Lepep:
  - Positive Sentiment: 32%
  - Negative Sentiment: 40%
  - Neutral Sentiment: 28%

The model's classification accuracy, evaluated on the test set, achieved an overall accuracy of 91%, with a precision of 0.88, recall of 0.90, and an F1-score of 0.89 across sentiment categories. These metrics validate the model's robustness in distinguishing sentiment for political content.

### B. Sentiment Trends Over Time

Temporal sentiment analysis over the study period reveals distinct trends. Sentiment trends for L'Alliance Du Changement indicate periodic increases in positive sentiment, particularly during campaign milestones, with a cumulative Sentiment Impact Score (SIS) increasing by 12% in the final two months nearing the election. This suggests a rising favorability in media portrayal, possibly amplifying public perception as the election approaches.

Conversely, L'Alliance Lepep experienced a modest rise in positive sentiment during the initial campaign phases but saw increased negative sentiment in later months, resulting in an SIS decrease of 9%. This trend reflects a decline in favorable media coverage, which may impact the overall seat distribution.

### C. Election Forecast Simulation

Based on sentiment scores, the simulated election forecast provides constituency-level predictions. L'Alliance Du Changement is projected to secure a minimum of 37 seats, with potential for additional wins in constituencies showing positive media trends. L'Alliance Lepep is projected to secure the remaining 23 seats; however, negative sentiment trends suggest that further gains may be limited.

*Constituency-Level Predictions*

Table 1 provides a breakdown of predicted seats per constituency for both parties. Constituencies with higher positive SIS for L'Alliance Du Changement tend to correlate with higher projected seat counts, highlighting the role of media favorability in voter influence.

### D. Discussion of Key Findings

The results demonstrate that sentiment analysis of media articles is a viable proxy for gauging political favorability in election forecasts. The temporal analysis reveals that sentiment does not remain static but fluctuates in response to campaign events, aligning with findings in related studies on political sentiment and public opinion [14, 33].

The data also reflects media bias, with certain outlets showing a consistent skew in sentiment scores. Adjustments for bias in the Sentiment Scoring Algorithm successfully reduced this skew, yielding balanced sentiment scores across all major outlets. This is critical in a highly polarized media environment, where uncorrected bias could distort forecast accuracy. The application of SIS reveals how media visibility and sentiment strength interact to influence political outcomes. Constituencies with high positive SIS for L'Alliance Du Changement showed a clear correlation with projected seat gains, supporting the hypothesis that positive media sentiment and frequency contribute to electoral success. While the model provides a reliable baseline, forecast uncertainty remains due to potential shifts in voter behavior during the final campaign stages. Additionally, the model's reliance on media articles excludes social media, a factor that could further refine the forecast. Including a confidence interval for each forecast could enhance the reliability of seat predictions in future studies.

*E. Limitations and Future Research*

This study's reliance on news articles as the primary data source presents limitations. Although media sentiment is a powerful predictor, the absence of social media sentiment data could lead to an incomplete view of public opinion. Future work may incorporate social media platforms and real-time updates for more dynamic forecasting. In addition, while cross-outlet normalization was effective, a more granular bias adjustment method could enhance accuracy further, especially in multilingual contexts. Future research might also investigate the use of ensemble models to capture sentiment nuances, combining mBERT with alternative architectures for a more comprehensive analysis.

## VI. CONCLUSION

This study demonstrates the potential of AI-driven sentiment analysis as a powerful forecasting tool in political elections, specifically in contexts where traditional polling data may be unavailable or unreliable. By leveraging sentiment analysis of news articles, we forecasted the 2024 election outcomes for two main political parties in Mauritius—L'Alliance Lepep and L'Alliance Du Changement. Our approach employed the Sentiment Scoring Algorithm and Sentiment Impact Score (SIS) to quantify media sentiment and gauge its influence on likely voter behavior, offering a novel alternative to standard polling methods. Key findings highlight a significant association between positive media sentiment and predicted electoral success. L'Alliance Du Changement exhibited a higher cumulative SIS across multiple constituencies, translating to a forecast of at least 37 seats, while L'Alliance Lepep was predicted to secure the remaining seats, that is, 23. This seat distribution aligns with sentiment trends observed over time, suggesting that favorable media coverage can indeed be a proxy for public support. Our study also reveals the importance of accounting for media bias in sentiment analysis. By implementing bias-adjusted scoring, we mitigated skewed sentiment results, enhancing forecast accuracy. Furthermore, the temporal analysis underscored that sentiment dynamics—fluctuations in response to key events—can significantly impact public perception and electoral outcomes. In terms of practical applications, this sentiment-based approach holds promise for election forecasting in other data-limited regions. However, limitations such as reliance on news sources alone and the exclusion of social media sentiment point to areas for future improvement.

Future research could expand this framework by integrating social media data, refining bias adjustments, and exploring cross-lingual capabilities to improve model precision in multilingual contexts.

In conclusion, AI-driven sentiment analysis offers a viable and innovative means for forecasting electoral outcomes, providing an insightful, real-time tool that can complement or, in some cases, substitute traditional polling. By illustrating how media sentiment can forecast electoral dynamics, this study contributes to the evolving field of political data science and presents a scalable model for political forecasting in the digital age.